\documentclass{cimento_arxiv}

\usepackage{bm}
\usepackage{graphicx}  
\usepackage{amsmath}
\DeclareMathOperator*{\argmax}{arg\,max}

\usepackage{fancyvrb}

\title{Training collective variables for enhanced sampling\\via neural networks based discriminant analysis}
\shorttitle{Training collective variables for enhanced sampling}

\author{Luigi Bonati \from{ins:x}\from{ins:y}\from{ins:z}}
\instlist{
    \inst{ins:x} Department of Physics, ETH Zurich, Switzerland
    \inst{ins:y} Institute of Computational Science, Universit\'a della Svizzera italiana Lugano, Switzerland
    \inst{ins:z} Atomistic Simulations, Italian Institute of Technology, Genova, Italy
    }

\begin{document}

\maketitle

\begin{abstract}
A popular way to accelerate the sampling of rare events in molecular dynamics simulations is to introduce a potential that increases the fluctuations of selected collective variables. For this strategy to be successful, it is critical to choose appropriate variables. Here we review some recent developments in the data-driven design of collective variables, with a focus on the combination of Fisher's discriminant analysis and neural networks. This approach allows to compress the fluctuations of metastable states into a low-dimensional representation. We illustrate through several examples the effectiveness of this method in accelerating the sampling, while also identifying the physical descriptors that undergo the most significant changes in the process.
\end{abstract}

\section{Introduction: rare events in molecular dynamics}
Atomistic simulations~\cite{ref:1_frenkel} play an important role in many fields of science.  
One of the prominent techniques is Molecular Dynamics (MD), where the time evolution of the system is obtained by integrating Newton's equation of motion. 
However, its direct application is hindered by the fact that many systems present high free energy barriers between metastable states, and consequently transitions occur on  timescales too long to be simulated~\cite{ref:2_valsson}. This is why such processes are called rare events. A variety of important phenomena fall into this category, including phase transitions and chemical reactions, as well as biological processes such as protein folding and ligand binding. 
 
\section{Accelerating the sampling with collective variables}
Several enhanced sampling methods have been devised to extend the scope of MD simulations. A large family of these techniques relies on the identification of a small set of order parameters $\bm{s}$, called collective variables (CVs), which are functions of the atomic positions $\bm{R}$~\cite{ref:2_valsson}. 
The potential energy surface $U(\bm{R})$ is modified by the addition of an external bias potential $V\left(\bm{s}(\bm{R})\right)$ which enhances the fluctuations of the selected variables by lowering the energetic barriers.
This increases the sampling of the transition region, in turn accelerating the occurrence of rare events. 

A popular method is metadynamics~\cite{ref:3_laio}, in which the bias potential $V(\bm{s})$ is expressed as a sum of repulsive Gaussians that are deposited at the points visited in the CVs space. In this way, the system is discouraged from re-exploring configurations that have already been visited and can escape local minima. A recent evolution of metadynamics, called on-the-fly probability enhanced sampling (OPES), builds the bias potential from an on-the-fly estimate of the probability distribution $P(\bm{s})$~\cite{ref:4_invernizzi}. 

\section{Data-driven identification of collective variables}
For these methods to be successful, the collective variables must be related to the slow relaxation modes of the system or, in other terms, to the transition pathways connecting local minima. Since this information is usually not known beforehand, the quest for such variables often becomes a chicken-and-egg problem. If we have ergodic simulations we can extract appropriate CVs based on free energy paths~\cite{ref:5_branduardi} or slowest relaxation modes~\cite{ref:6_hernandez}, but for simulations to be ergodic we typically need knowledge of good CVs.

Here we look at the problem from a different perspective, and review some recent developments which exploits machine learning (ML) techniques to design CVs using information limited to the metastable states. ML methods have risen to prominence in recent years, and there is no shortage of applications to atomistic simulations~\cite{ref:7_riniker,ref:7_noe,ref:10b_bonati} and enhanced sampling~\cite{ref:8_ferguson,ref:9_tiwary,ref:10_bonati}. Both supervised~\cite{ref:11_pande} and unsupervised methods~\cite{ref:12_ferguson,ref:12_tum} have been applied to the search of data-driven CVs. We discuss here in detail the approaches that exploit Fisher's linear discriminant analysis~\cite{ref:13_mendels,ref:13b_piccini}, and in particular recent advances in which this statistical method has been coupled with neural networks~\cite{ref:14_bonati}.

\subsection{Linear Discriminant Analysis}
Linear Discriminant Analysis (LDA)~\cite{ref:15_welling} is a supervised method, first developed by Fisher, which is routinely used to perform classification and dimensionality reduction tasks. In summary, LDA seeks for a linear projection of the input features into a lower-dimensional space such that the classes are maximally separated. 
We illustrate here for simplicity the case of two classes, \textit{i.e.}~two metastable states $A$ and $B$. First, we choose a set of descriptors $\bm{d}$ that can be, for instance, distances, angles, dihedrals or coordination numbers. We run short MD simulations in basin A and B and compute the values of such descriptors.  
We use the average values $\bm{\mu}_A$, $\bm{\mu}_B$ and the variance matrices $\bm{\Sigma}_A$, $\bm{\Sigma}_B$ of the descriptors to characterize the basins and define the within-scatter matrix as $\bm{S}_w=\bm{S}_A+\bm{S}_B$ and the between-scatter matrix as $\bm{S}_b=(\bm{\mu}_A-\bm{\mu}_B)(\bm{\mu}_A-\bm{\mu}_B)^T$.

Fisher's criterion states that the linear combination $\bar{\bm{w}}$ that optimally separates classes is the one for which the ratio of the two scattering matrices in the projected space is maximal:
\begin{equation}
\label{eq:fisher}
    \bar{\bm{w}} = \argmax_{\bm{w}} \frac{\bm{w}^T \bm{S}_b\, \bm{w}}{\bm{w}^T \bm{S}_w\,\bm{w}}
\end{equation}

From a practical perspective, we can recast the above task as a generalized eigenvalue problem: $\bm{S}_b \bar{\bm{w}} = u\, \bm{S}_w\bar{\bm{w}}$, where the eigenvalue $u$ measures the degree of separation of the two states along the corresponding eigenvector $\bar{\bm{w}}$.
Once $\bar{\bm{w}}$ is found, the one-dimensional projection $s=\bar{\bm{w}}^T\bm{d}$ can be used as CV. A modified version of this procedure has been proposed to better describe the fluctuations of metastable states, which replaces the arithmetic mean in the definition of $\bm{S}_w$ with an harmonic one~\cite{ref:13_mendels}. 

\begin{figure}[t]
\label{fig:scheme}
    \begin{center}
        \includegraphics[width=0.7\columnwidth]{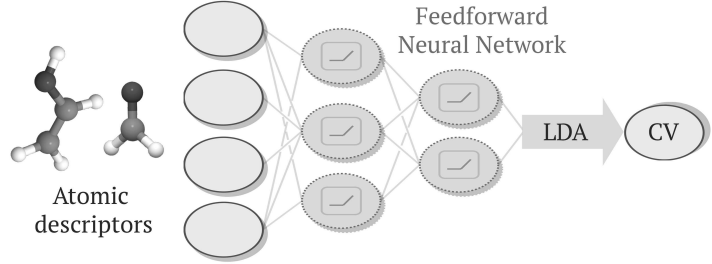}   
        \caption{Deep-LDA CV scheme. The NN performs a nonlinear trasformation of the input descriptors into to a compressed latent space, where LDA is applied.}
    \end{center}
\end{figure}

\subsection{Neural network based discriminant analysis}

While LDA provides a robust method for finding low-dimensional representations, it suffers from several limitations, the most important being that it can only provide a linear combination of the given descriptors. This precludes the application of the LDA CV to complex systems, where several descriptors need to be combined in a nonlinear way to fully capture the dynamics of the process. For this reason, we proposed to extend it with neural networks (NNs), similarly to what has been done in the context of computer science for classification~\cite{ref:16_dorfer}. This is achieved by using a NN to perform a nonlinear featurization of the input descriptors, that is optimized using Fisher's linear discriminant as the objective function. In the following we refer to this method as Deep-LDA.

\begin{figure}[b!]
\label{fig:lda-deeplda}
    \begin{center}
        \includegraphics[width=0.95\columnwidth]{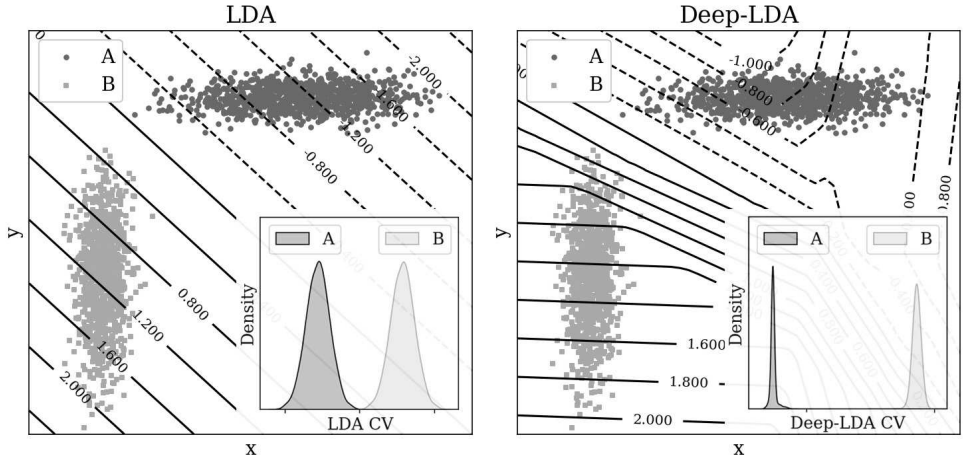}   
        \caption{Comparison of LDA (left) and Deep-LDA (right) for a 2D Gaussian toy model. The points represent the two states, while the lines correspond to the CV isolines. In the case of LDA, the CV projection, orthogonal to the isolines, is a straight line, while for Deep-LDA it connects the fluctuations of the two states in a nonlinear fashion. In the insets we report the density distribution along the CVs, to highlight the different discriminative power.}
    \end{center}
\end{figure}

As shown in Fig.~1, the descriptors are fed into a NN that compresses the information into a lower-dimensional space. LDA is then performed in this latent space, and the eigenvalue $u$ of the generalized eigenvalue equation is used as loss function to maximize the discriminative power of the NN. By applying Cholesky decomposition we can rewrite the generalized eigenvalue equation into a standard one. This allows to train the NN via gradient descent using the machine learning library PyTorch \cite{ref:pytorch}. Furthermore, a few regularizations are proposed to make the optimization more stable and to provide bounds to the objective function~\cite{ref:14_bonati}.

In fig.~2 we exemplify the behavior of LDA and Deep-LDA in the simple case of two-dimensional Gaussians. We observe that, due to the nonlinear transformation performed by the NN, the Deep-LDA CV discriminates the two states better than its linear version.
\section{\textit{Intermezzo}: how to train and apply the Deep-LDA CV}
Before moving on to applications, we summarize the main steps of this method from a practical viewpoint. Once a set of descriptors has been chosen, the first step is to simulate short MD trajectories in the metastable states and compute the descriptors' values. The NN can be trained on these datasets using the Pytorch library, following the tutorial on Github~\cite{ref:github}. 
Finally, the CV can be used along with enhanced sampling methods through the open-source library PLUMED\cite{ref:plumed}, taking advantage of an interface we developed to import Pytorch models. Codes and input files are available also on the PLUMED-NEST~\cite{ref:nest} repository with \texttt{plumID:20.004}. Below is an example of a PLUMED input file, which shows the ease of use of this method.

\begin{figure}[h!]
    \begin{center}
    \begin{BVerbatim}[fontsize=\small,commandchars=\\\{\},baselinestretch=1.2]
--------------------- \textbf{PLUMED INPUT} ---------------------
 \fbox{COMPUTE DESCRIPTORS}
   d1: DISTANCE ATOMS=1,2 ...
   dN: DISTANCE ATOMS=17,19
 \fbox{LOAD DEEP-LDA CV}		
   deep: PYTORCH_MODEL MODEL=model.pt ARG=d1,...,dN
 \fbox{APPLY BIAS POTENTIAL}		
   opes: OPES_METAD ARG=deep.node-0 PACE=500 BARRIER=40
--------------------------------------------------------
    \end{BVerbatim}
    \end{center}
\end{figure}
\vspace{-8mm}

\section{Applications}

\subsection{Enhancing the sampling with Deep-LDA}
As an application of the Deep-LDA CV, we examine the chemical reaction between vinyl alcohol and formaldehyde which has been studied in ref.~\cite{ref:14_bonati}. Since the free energy barrier between reactants and products is on the order of 170 kJ/mol (68 $k_B T$ at room temperature), it is impractical to observe any transition in a MD simulation without resorting to enhanced sampling methods. 

The set of descriptors consists of all 40 interatomic distances, whose values are measured in short simulations (5 ps) of the reactants and products. The Deep-LDA CV is trained on these datasets, using a NN with 3 hidden layers and \{24, 12, 4\} nodes per layer. Further computational details can be found in ref.~\cite{ref:14_bonati}.
The CV is used to accelerate sampling in combination with the OPES method, which leads to the observation of several transitions between the two states. This allows the free energy profile of the reaction to be computed with great accuracy (Fig.~3a). This sampling efficiency implies that the CV, even if based only on the configurations of the metastable states, is able to capture the concerted reaction mechanism. 

\begin{figure}[t!]
\label{fig:aldol}
    \begin{center}
        \includegraphics[width=\columnwidth]{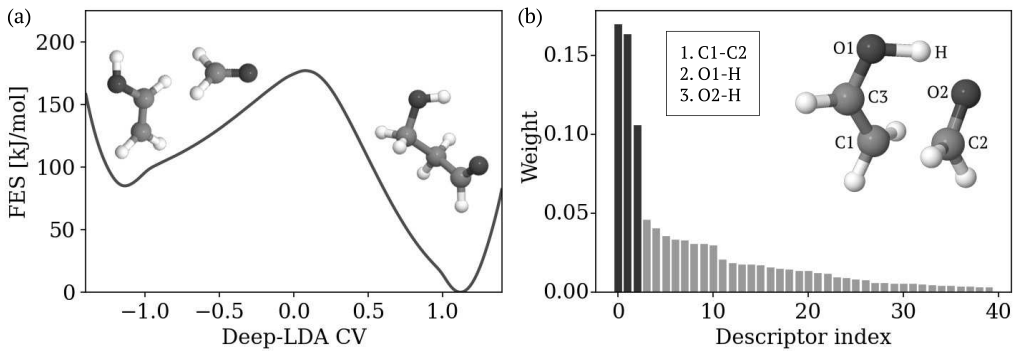}   
        \caption{(a) Free energy as a function of the Deep-LDA CV, obtained from the OPES simulation via a reweighting procedure. The error, calculated with a block average, is about 1 kJ/mol. Snapshots of the reactants (left) and product (right) are also displayed. (b) Contributions of the descriptors to the CV. They are computed as the sum of the modulus of the weight parameters between the input and the first hidden layer, multiplied by the standard deviation of the inputs. The weights are normalized so that their sum equals one. The three dominant descriptors are colored in  black, and the corresponding distances are reported in the box.}
    \end{center}
\end{figure}

\subsection{Extracting physical insights from the CVs}
A notable feature of these data-driven CVs is their ability to highlight the most important descriptors, thereby deepening the physical understanding of the process. We illustrate this property with two examples. 

The first is the chemical reaction described above: in Fig.~3b the descriptors are ranked according to their contribution to the CV, from which we can recognize three dominant distances.
This fact is in accordance with the chemical changes that the system undergoes during the reaction, in which a bond is formed between the two carbon atoms and, simultaneously, a proton transfer occurs between the oxygen atoms. 

The second example is the study of the binding between a calixarene host and a set of ligands, reported in \cite{ref:17_rizzi}. The Deep-LDA CV is used to include the effect of water in the study of the binding process, which is otherwise typically modeled using information limited to the ligand and the host. The analysis of the descriptors' weights allows to understand the role of water in the binding and unbinding events, and the inclusion of this contribution in the CV set results in a very efficient sampling of the process \cite{ref:17_rizzi}.

\section{Conclusions}

The combination of Fisher's linear discriminant and neural networks allows the design of data-driven collective variables using information limited to local minima. This approach has proven effective in accelerating the sampling of different types of rare events and identifying the most relevant atomic descriptors. 
It represents a promising tool for studying complex systems, and can be used as an initial guess for methods designed to extract CVs from enhanced sampling simulations~\cite{ref:18_tica,ref:19_pande}. 

\acknowledgments
The author thanks Dr. Michele Invernizzi for carefully reading the manuscript. The research was supported by the NCCR MARVEL, funded by the Swiss National Science Foundation, and European Union Grant No.~ERC-2014-AdG-670227/VARMET.

\end{document}